\documentclass[12pt,a4paper]{book}
\usepackage{epsf}
\usepackage[small,bf]{caption}
\begin{document}

\begin{center}

\thispagestyle{empty}

\phantom{.}

\vspace{4cm}

{\huge \bf Cosmic Dust} 

\vspace{0.5cm}

{\large \bf The Ulysses perspective}

\vspace{3cm}

{\em
        Eberhard Gr\"un$^1$,
Harald Kr\"uger\,$^1$, Markus Landgraf\,$^{2}$}

\bigskip
\bigskip

{\small $1$: Max-Planck-Institut f\"ur Kernphysik, Postfach 10\,39\,80, 69029 Heidelberg,
    Germany \\
        $2$: ESA-ESOC, Robert-Bosch-Str. 5, 64293 Darmstadt, Germany}

\vspace{4cm}

To appear in: {\bf The heliosphere at solar minimum: The Ulysses
perspective}, \\ 
eds. A. Balogh, R. Marsden, E. Smith; Springer Praxis, 2001

\end{center}

\vspace{5cm}
\thispagestyle{empty}

\chapter{Cosmic Dust}

\bigskip

By {\em
        Eberhard Gr\"un$^1$,
Harald Kr\"uger\,$^1$, Markus Landgraf\,$^{2}$}

{\small $1$: Max-Planck-Institut f\"ur Kernphysik, Postfach 10\,39\,80, 69029 Heidelberg,
    Germany \\
        $2$: ESA-ESOC, Robert-Bosch-Str. 5, 64293 Darmstadt, Germany}

\section{Introduction}

There are several methods to study cosmic dust. The traditional method
to determine the global structure of the interplanetary  dust cloud is
by zodiacal light observations (for a review see, e.g. 
\cite{leinert-and-gruen-1990}). Before Ulysses, zodiacal light
observations have provided the primary means of studying the
out-of-ecliptic  distribution of the interplanetary dust cloud. By
inversion of these scattered light observations it has been derived 
that the grain 
properties depend upon their elevation above the ecliptic  plane, i.e.
upon the inclination of their orbits \cite{levasseur-regourd-1991}. Several
3-dimensional models of the intensity,  polarization and color of
the zodiacal dust cloud have been developed 
\cite{giese-et-al-1985,giese-et-al-1986}.

Thermal infrared observations by IRAS  
\cite{hauser-et-al-1984} and especially
COBE  \cite{reach-et-al-1995} give further constraints on the
zodiacal  dust cloud outside and above the Earth's orbit.
Interpretations of these  observations, however, rely heavily on
assumptions about the size distribution  and material properties of
the dust particles.  More recently, the infrared spectral energy
distribution and brightness  fluctuations in the zodiacal cloud have
been studied with ISO
\cite{abraham-et-al-1999-a,abraham-et-al-1999-b}. These observations
will also allow for a better determination of the size distribution of
the  interplanetary dust grains.

The dust experiments on board the Pioneer~10 and  Pioneer~11 spacecraft
provided information on the radial dependence of the spatial density
of large dust grains ($ \rm \sim 20 \,\mu m$ diameter) outside Earth's
orbit. Between 1 and 3.3~AU  these experiments detected a decrease of
the dust abundance proportional to $ r^{-1.5}$, $r$ being the 
heliocentric distance \cite{humes-et-al-1974,hanner-et-al-1976}.
The photometer on board Pioneer~10 did not record any scattered  sunlight
above the background beyond the asteroid belt  while the penetration
experiment recorded dust impacts  out to about 18 AU heliocentric distance 
at an almost  constant rate \cite{humes-1980}. 

Measurements of dust in the inner solar system  with the Pioneer~8, Pioneer~9
and Helios space probes and the  HEOS 2 satellite showed that there
are several populations of dust particles which possess different
dynamical  properties. A population of slow-moving, small 
($ 10^{-16}\, {\rm kg } < m < 10^{-14}\, {\rm kg}$) particles has been observed 
by the Pioneer  8/9 \cite{berg-and-gruen-1973} and HEOS 2 
\cite{hoffmann-et-al-1975} dust experiments. The
Helios dust experiment  \cite{gruen-et-al-1980} confirmed these particles
which orbit  the Sun on low eccentric orbits (e$<$0.4). These low 
angular momentum ``apex'' particles were thought to
originate from collisional break-up of  larger meteoroids in the inner
solar system \cite{gruen-and-zook-1980}. Another population which consists of very 
small particles ($ m \sim 10^{-16} {\rm kg}$) has been detected by the 
Pioneer 8/9 and  Helios space probes arriving at the sensors from 
approximately the solar direction.
These particles have been identified 
\cite{zook-and-berg-1975} as small
grains generated in the inner solar  system which leave the solar
system on hyperbolic orbits  due to the dominating effect of the
radiation pressure  force ($\beta$-meteoroids). 

The orbital distribution of larger meteoroids  is best known from
meteor observations. 
Sporadic  meteoroids move on orbits with an average eccentricity  of
0.4 and an average semi-major axis of 1.25 AU at  Earth's orbit
\cite{sekanina-and-southworth-1975}.
Helios measurements allowed the identification of an interplanetary dust
population \cite{gruen-et-al-1980} which consists of particles on highly
eccentric orbits  (e $>$ 0.4) and with semi-major axes larger than 0.5~AU. 
Pioneer~11 in-situ data obtained between 4 and 5~AU are best explained by
meteoroids moving on highly eccentric orbits  \cite{humes-1980}. These
particle populations resemble most  closely the sporadic meteor
population.

Before Ulysses, dust has been observed in the magnetosphere of Jupiter, both  
by in-situ experiments on board Pioneer~10 and Pioneer~11, and by  remote 
sensing instrumentation on board Voyager~1 and Voyager~2.  Pioneer~10/11 
measured a 1000 times higher flux of  micrometeoroids in the vicinity of
Jupiter compared with  the interplanetary flux. The Voyager imaging
experiment  detected a ring of particulates at distances out to 2.5 
jovian radii, as well as volcanic activity on the jovian satellite Io. 

One of the most important results of the Ulysses mission is the 
identification and characterization of a wide range of interstellar 
phenomena inside the solar system. A surprise was the identification
and  the dominance of the interstellar dust flux in the outer solar
system.  Before this discovery by Ulysses it was believed that
interstellar grains  are prevented from reaching the planetary region
by electromagnetic  interaction with the solar wind magnetic field.
The interplanetary zodiacal  dust flux was thought to dominate the
near-ecliptic planetary region while  at high ecliptic latitudes only
a very low flux of dust released from  long-period comets should be
present. Therefore, the characterization of  the interplanetary dust
cloud was the prime goal of the Ulysses dust investigation. 

Around Jupiter fly-by in 1992 Ulysses has discovered intense collimated 
streams of dust particles. The grains were detected at a distance out to 
2~AU from Jupiter along the ecliptic plane and at high ecliptic latitudes
\cite{gruen-et-al-1993}. Modelling of the dust trajectories showed that
the grains must have been ejected from the jovian system 
\cite{zook-et-al-1996}. Later measurements by the Galileo spacecraft within 
the jovian magnetosphere revealed Io as the source of the grains 
\cite{graps-et-al-2000}. 

\begin{figure}[tbh]
\vspace{-9.0cm}
\epsfxsize=13.2cm
\epsfbox{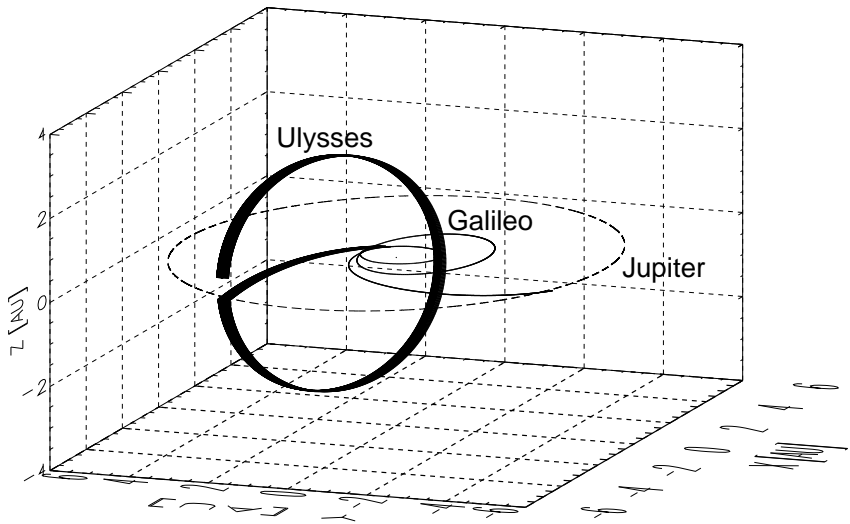}
\vspace{-1.0cm}
        \caption{\label{orbits}
Trajectories of the Ulysses (heavy solid line) and the Galileo (thin 
solid line) spacecraft. The Sun is in the center, Jupiter's (dashed) 
and Galileo's trajectories are in the ecliptic plane.
% (shaded). 
The initial trajectory of Ulysses from Earth to Jupiter
was also in the ecliptic plane. Subsequently, Ulysses was thrown into an
orbit plane inclined by $79^{\circ}$ to the ecliptic plane. Vernal 
equinox is towards the positive x direction and north is at the top. The
Ulysses trajectory is shown until the end of 1997.
}
\end{figure}

Here, we focus on measurements by the in situ dust detectors on board
the Ulysses and Galileo spacecraft to obtain  information on the large
scale structure and the dynamics of the interplanetary  dust cloud 
(Section~\ref{interplanetary_dust}),
dust in the environment of Jupiter (Section~\ref{jupiter_dust}), and 
interstellar dust (Section~\ref{interstellar_dust}). The dust 
detectors measure dust along the the spacecraft's interplanetary
trajectories. From these measurements  3-dimensional model of the
spatial dust distribution can be constructed. The  orbits of the Galileo
and Ulysses spacecraft are displayed in Fig.~\ref{orbits}.

\subsection{Dust objectives}

The Ulysses spacecraft was launched 
in October 1990 onto a direct trajectory  to Jupiter. After Jupiter
fly-by in February 1992, Ulysses was  thrown onto an orbit of $\rm
79^{\circ}$ inclination that passed close to the  ecliptic poles. The
passage from the south to the north pole took one year from  September
1994 to September 1995 and the passage through the ecliptic plane 
occurred in March 1995. Ulysses completed a full polar orbit in April
1998  when it reached its aphelion distance at 5.4 AU again, however,
this time Jupiter   was on the opposite side of the Sun.

Ulysses carried a high sensitivity dust detector to the outer  solar
system for the first time. The detector is five orders of magnitude more sensitive
than those  on the Pioneer 10~and Pioneer~11 spacecraft \cite{humes-1980}. The
objectives of the  Ulysses dust experiment  (as stated in the original
proposal \cite{gruen-et-al-1992-b}) were to 

\begin{itemize}
\item[(1)] determine the 3-dimensional structure of the zodiacal
cloud, 
\item[(2)] characterize its dynamical state, and 
\item[(3)] search for interstellar dust penetrating the solar system. 
\end{itemize}

The prime objective was the determination of the 3-dimensional dust
distribution  in the solar system. In order to reach this goal with
the help of out-of-ecliptic  Ulysses measurements, reference dust
measurements had to be provided in the  ecliptic plane. This was
almost ideally achieved by the Galileo mission that  was launched a
year earlier but took dust measurements for 6 years in  interplanetary
space near the ecliptic plane with a twin of the Ulysses  dust detector.
While Ulysses data are very good at determining the absolute  latitude
dependence (and inclination dependence) of meteoroids, measurements 
by Galileo near the ecliptic plane determine the radial and
eccentricity dependence  for low inclination orbits. The achievement
of the third Ulysses objective,  namely, the identification of
interstellar dust particles, required that  interstellar dust had to
be identified and separated from interplanetary dust.  For
measurements by both the Galileo and Ulysses detectors, this
distinction was  easy outside about 3 AU because it was found that the
flux of interstellar dust  grains dominated and differed significantly
from  prograde interplanetary dust, both in direction and speed 
\cite{gruen-et-al-1993, gruen-et-al-1994-b, baguhl-et-al-1995-a,
baguhl-et-al-1995-b}.

\section{Instrumentation}

Ulysses is a spinning spacecraft with its spin-axis being coincident with the
antenna pointing direction. Its antenna usually points towards Earth.
The Ulysses dust detector
is mounted at an angle of $\rm 85^{\circ}$ from the antenna direction.
The field-of-view (FOV) of the dust detector is a cone of
$\rm 140^{\circ}$ full angle and the spin averaged effective sensor area for
impacts varies with the angle between the impact direction and the
antenna direction
\cite{gruen-et-al-1992-b}.
%(cf. Gr\"un et al., 1992b).
Because the detector
is mounted almost perpendicular to the spacecraft spin-axis, the maximum
spin-averaged sensitive area of $\rm 200\,cm^2$ is reached for impacts along
a plane almost perpendicular to the antenna direction (the geometry of dust
detection during Jupiter fly-by is sketched in Fig.~\ref{geometry}).
The impact direction (rotation angle, ROT) in this plane
%perpendicular to the spacecraft axis
is determined by the spin position of the spacecraft around its spin-axis
at the time of a dust impact. The rotation angle is
zero when the dust sensor axis is
closest to the ecliptic north direction. The rotation angle is measured
in a right-handed system around the antenna direction.

\begin{figure}[ht]
\parbox{0.49\hsize}
{
\epsfxsize=7.6cm
\epsfbox{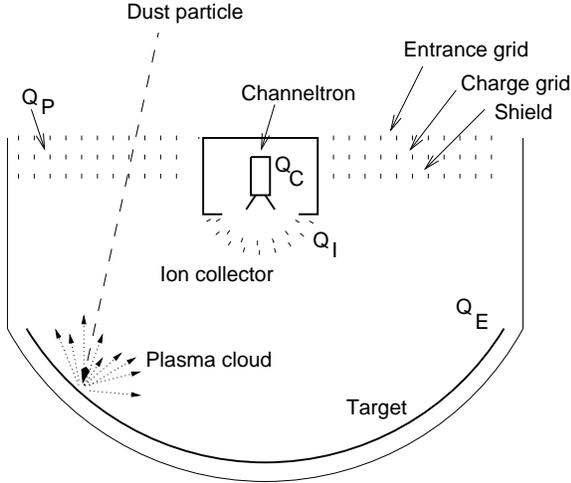}
}
\parbox{0.5\hsize}
{
        \caption{\label{ddsconf}
Schematic configuration of the Ulysses dust detector (GRU). Particles
hitting the target create a plasma cloud and
up to three charge signals ($ Q_{\rm I}, Q_{\rm E}, Q_{\rm C}$) are used for dust impact
identification.
}
}
\end{figure}

The Ulysses dust detector
\cite{gruen-et-al-1992-b}
%(cf. Gr\"un et al., 1992b).
is an impact
ionization sensor which measures the plasma cloud generated upon impact
of submicrometer- and micrometer-sized dust particles onto the detector target
(cf. Fig.~\ref{ddsconf}). Up to 3 independent measurements
of the ionization cloud created during impact are used to derive both the
mass and the impact speed of the dust grains \cite{gruen-et-al-1995-a}.
%(Gr\"un et al., 1995).
The detector mass threshold, $ m_{\rm t}$, is proportional to the positive charge
component, $ Q_{\rm I}$, of the plasma produced during the impact, which itself
strongly depends on the impact speed, $v$. Using the calibration parameters
$( Q_{\rm I}/m)_0$, $ m_0$, $ v_0$, and $\alpha$, which have been
approximated from detector calibrations
%(Gr\"un et al., 1995a),
\cite{gruen-et-al-1995-a}, the
corresponding mass threshold can be calculated:
\begin{equation}
    m_{\rm t} = \frac{Q_{\rm I}}{(Q_{\rm I}/m)_0} = m_0\left (\frac{v_0}{v}\right )^{\alpha},
\end{equation}
with $\alpha \sim 3.5$. For example, an impact charge of
$ Q_{\rm I} = \rm 8\cdot 10^{-14}\, C$ refers to a mass threshold
$  m_{\rm t} = \rm 3\cdot 10^{-17}\, kg$ at $ 20\, \rm km\,s^{-1}$ impact speed.

The dynamic range of the impact charge measurement is
$10^6$ which is also the dynamic range of the mass determination for particles with
constant impact speeds. The calibrated speed range of the
instrument is ${\rm 2\,km\,s^{-1}} \leq v \leq {\rm 70\,km\,s^{-1}}$ which
corresponds to a calibrated mass range of $\rm 10^{-19 \ldots -9}\,kg$.
Impact speeds can be determined with an accuracy of about a factor two and
the accuracy of the mass determination of a single particle is about a factor of 10.

Impact-related data such as up to three charge measurements, impact time and
rotation angle are normally all transmitted to Earth for each impact. Thus,
a complete record of impact charge, particle mass, impact velocity,
impact time and impact direction is available for each detected particle.
Impact rates are derived from the number of detections within a given time
interval.

In addition to Ulysses, the Galileo spacecraft carries a nearly identical
dust detector on board. After being launched in 1989 Galileo
traversed interplanetary space
and was injected into a bound orbit about Jupiter in 1995.
Galileo dust data supplement Ulysses measurements because Galileo measured
dust along the ecliptic plane at times when Ulysses was at high
ecliptic latitudes. This allowed for the determination of radial and
latitudinal variations of the interplanetary dust cloud
\cite{gruen-et-al-1997-a}.
%(Gr\"un et al., 1997a).
Dust data obtained with the detectors on board both spacecraft -- Ulysses and
Galileo -- can be found in the literature
\cite{gruen-et-al-1995-b,gruen-et-al-1995-c,krueger-et-al-1999-b,
krueger-et-al-1999-c,krueger-et-al-2001a,krueger-et-al-2001b}

\section{Interplanetary dust background in the ecliptic plane and above the solar poles}
\markright{INTERPLANETARY DUST}
\label{interplanetary_dust}

\begin{figure}[tbh]
\vspace{-1.7cm}
\epsfxsize=11.2cm
\epsfbox{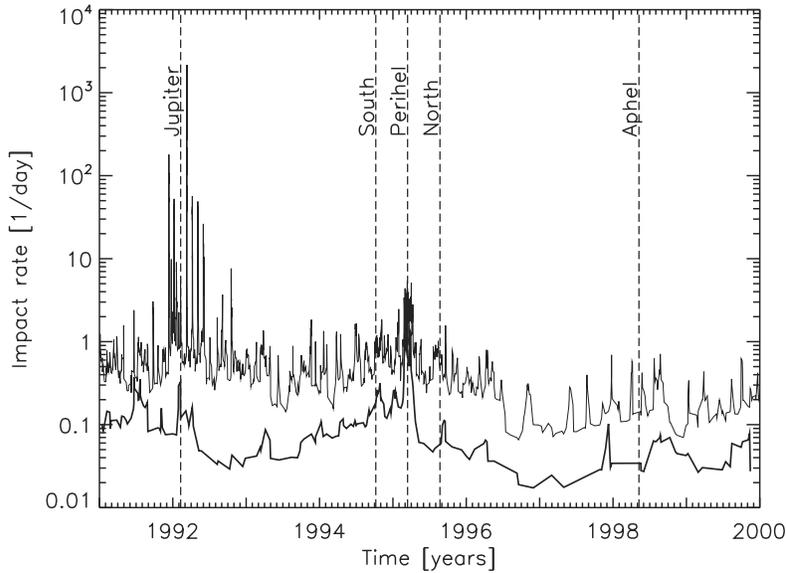}
\vspace{-4.5cm}
        \caption{\label{Fig3_2_1}
Dust impact rates observed by Ulysses. The impact rates shown
cover the periods from launch to Jupiter fly-by and the full out-of-ecliptic
orbit from Jupiter over the poles of the Sun, through the ecliptic plane
(perihelion) and out to aphelion at Jupiter distance. The impact rates are all
impacts recorded (upper trace) and impacts of dust particles with
$ m >\rm 10^{-15}\,kg$ (lower trace). The impact rates are sliding
means always including six impacts.
}
\end{figure}

The Ulysses dust measurements provide -- as a function of time and spaceprobe
position -- the flux of particles as a function of impact direction, speed, and
particle mass. Here we include data obtained by Ulysses from launch
to the completion of the first inclined orbit. Shortly after launch the impact
rate of $ \rm 10^{-17}\, kg$ and smaller particles declined from a few impacts
per day to about one impact per 3 days (Fig.~\ref{Fig3_2_1}). Most of the time of
interplanetary cruise until 1996 the dust impact rate stayed at this level.
In late 1996 the impact rate decreased again by about a factor of 3. For about one
year around Jupiter fly-by the impact rate of the smallest impacts increased
by up to a factor of 1000 for periods of one to several days. These dust
streams are the second discovery of the Ulysses dust instrument
\cite{gruen-et-al-1993}.
Their characteristics will be discussed
in Section~\ref{jupiter_dust}. 

At high ecliptic latitudes
close to the solar poles a surprisingly large dust flux was recorded. Around
the time of ecliptic plane crossing the impact rate of big particles
($m \rm > 10^{-15}\,kg$)
increased by about a factor of 10, whereas less massive particles showed a smaller
enhancement. Particles smaller than $\rm 10^{-17}\,kg$ were recorded in abundance
for some time shortly after launch,
around Jupiter fly-by as jovian dust streams,
%during all times of Jupiter dust streams,
and around the times of the polar passages. These events will be discussed in the
sections on $\beta$-meteoroids (Section~\ref{beta_meteoroids}), and on Jupiter
dust streams (Section~\ref{jupiter_dust}).

\begin{figure}[htb]
\vspace{-4.6cm}
\epsfxsize=11.8cm
\epsfbox{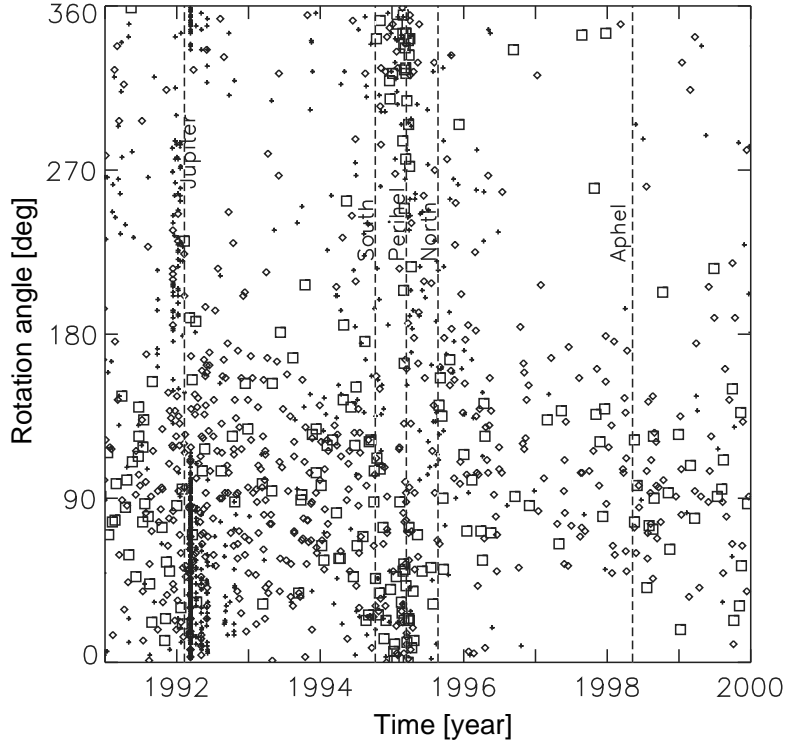}
\vspace{-0.4cm}
        \caption{\label{Fig3_2_2}
Dust impact directions (rotation angle) observed by Ulysses.
The rotation angle is defined in Fig.~\ref{geometry}.
The rotation angle is $0^{\circ}$ when the
dust sensor axis points closest to the ecliptic north pole. At
rotation angle $90^{\circ}$ the detector axis points
parallel to the ecliptic plane in the direction of
planetary motion, and at $180^{\circ}$ it points closest to
the south ecliptic pole.  The data
cover the periods from launch to Jupiter and a full out-of-ecliptic orbit.
The smallest impacts of particles with masses $m \rm \leq 10^{-17}\,kg$
are denoted by crosses, medium massive particles
($\rm  10^{-17}\,kg < \it m \leq \rm 10^{-15}\,kg$) are denoted by diamonds,
and most massive particles ($m \rm > 10^{-15}\,kg$) are denoted by squares.
}
\end{figure}

The directional distribution of dust impacts (Fig.~\ref{Fig3_2_2}) shows a similar
separation in different categories as the impact rate. Here, we concentrate on
particles bigger than $\rm 10^{-17}\,kg$, smaller particles are discussed in their 
respective sections. The impact directions were concentrated at rotation angles
of about $\rm 90^{\circ}$ for most of the time. This direction is compatible with the flow
direction of interstellar grains through the solar system
(cf. Section~\ref{interstellar_dust}). Only around
the time of ecliptic plane crossing the mean impact direction  shifted to about
$\rm 0^{\circ}$ rotation angle. This direction is compatible with the flux of
interplanetary meteoroids around the Sun. Mostly during the passage from the north
pole to Ulysses' aphelion also a few  big impacts were recorded from about
$\rm 270^{\circ}$ rotation angle, i.e. the direction one would expect particles
on prograde bound orbits to arrive from.

\subsection{Ulysses' South-North traverse}

The Ulysses mission is especially well suited to obtain a latitudinal profile of
the interplanetary dust cloud. In the distance range from 2.3 to 1.3 AU,
Ulysses passed from close to the ecliptic south pole ($\rm -79^{\circ}$ ecliptic
latitude) through the ecliptic plane to the north pole ($\rm +79^{\circ}$). The
outer portions of the out-of-ecliptic orbit (beyond 2.3 AU) are not suited for
characterizing interplanetary dust because of the dominant interstellar dust
population \cite{baguhl-et-al-1995-a}.%(Baguhl et al., 1995a).

\begin{figure}[ht]
\epsfxsize=8.2cm
\epsfbox{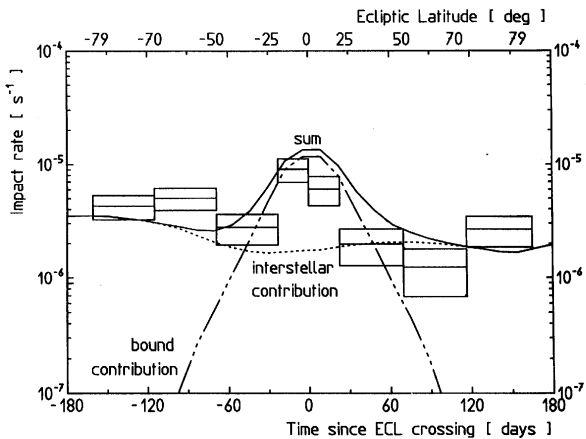}
        \caption{\label{Fig3_2_3}
Dust impact rate (impact charge $Q_{\rm I} \rm  > 8 \cdot 10^{-14}\,C$) observed by
Ulysses during its South-North traverse around the time of ecliptic
plane crossing (ECL). The top scale gives the spacecraft latitude. The
boxes indicate mean impact rates and standard deviations. The observed
impact rates are compared with models of the two major dust components
in the solar system: interstellar dust and interplanetary dust on bound
orbits about the Sun.
}
\end{figure}

The relation between the dust density at a given latitude and the inclination
distribution is simply given by the fact that dust particles recorded at ecliptic
latitude $\lambda$ must have inclinations $ i \rm \geq \lambda$ in order to reach
this latitude. Figure~\ref{Fig3_2_3} shows the impact rate during the pole-to-pole traverse.
The passages over the solar south and north poles occurred 170 days before and after
ecliptic plane crossing, respectively. A total of 109 impacts (with impact
charges $Q_{\rm I} > \rm 8\cdot 10^{-14}\,C$) were recorded during this time. The impact rate
stayed relatively constant except for the maximum ($\rm 9\cdot 10^{-6}\,s^{-1}$) during 
ecliptic plane crossing. The impact rate at the northern leg is about a factor of 2
below that of the southern one. This is due to the varying spacecraft attitude which
followed the direction to the Earth. This variation of spacecraft attitude is also
reflected in the variations of rotation angles of the impacts which were detected
during the S-N traverse. Over the south solar pole most large impacts
(with impact charges $Q_{\rm I} \rm > 8\cdot 10^{-14}\,C$) occurred at rotation angles
between $0^{\circ}$ and $150^{\circ}$ which includes the interstellar direction
(cf. Fig.~\ref{Fig3_2_5}).
Closer to the ecliptic plane the rotation angle range of large impacts widened and
moved further to the north direction (rotation angle $ \sim 0^{\circ}$). At ecliptic
plane passage these impacts were recorded in a wide range around the north direction
($0^{\circ}$ to $100^{\circ}$ and $200^{\circ}$ to $360^{\circ}$). On the northern
pass, rotation angles covered the whole range and above the north pole it ranged
from $50^{\circ}$ to $200^{\circ}$, again including the interstellar direction.

\begin{figure}[ht]
\vspace{-1.3cm}
\parbox{0.6\hsize}
{
\epsfxsize=9.9cm
\epsfbox{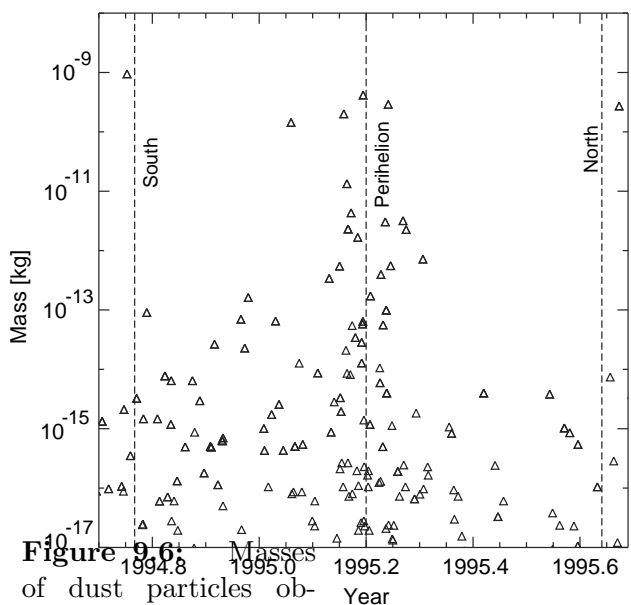}
\vspace{-5.0cm}
}
\parbox{0.35\hsize}
{
        \caption{\label{Fig3_2_4}
Masses of dust particles observed by Ulysses during its
South-North traverse. Ecliptic plane crossing coincided with perihelion
passage.
}
}
\end{figure}

Figure~\ref{Fig3_2_4} shows the masses of dust particles detected during the S-N traverse.
20 particles with masses greater than $\rm 10^{-13}\,kg$ were detected during the S-N
traverse, 15 of these particles were recorded during the 80 days when the spacecraft
was close to the ecliptic plane ($-30^{\circ} < \lambda < 30^{\circ}$). This increased
flux of big particles
near the ecliptic plane is obviously due to the zodiacal dust population. The mass distribution
during the S-N traverse seems to be composed of two distinct components: the
interstellar dust component, which has peak masses between $\rm 10^{-14}$ and
$\rm 10^{-15}\,kg$ and a bigger interplanetary dust component dominating the dust flux
in the near-ecliptic region.

\subsection{$\beta$-meteoroids}

\label{beta_meteoroids}

$\beta$-meteoroids are particles that leave the solar system on hyperbolic orbits
because of radiation pressure and electromagnetic solar wind interactions 
\cite{hamilton-et-al-1996}. They could only be observed by Ulysses for a few short 
periods during the Ulysses mission: the 
early mission phase, both polar passes 
\cite{baguhl-et-al-1995-a}, and the perihelion ecliptic plane crossing. 
Wehry and Mann \cite{wehry-and-mann-1999} have
identified $\beta$-meteoroids in the Ulysses data set and discuss some properties. 

\subsection{Mass distribution of interplanetary particles}

Although Galileo and Ulysses dust data cover a mass range
${\rm 10^{-19}\,kg} \leq m \leq {\rm 10^{-9}\,kg}$, the statistically best and most
complete measurements
range only from ${\rm 10^{-17}\,kg} \leq m \leq {\rm 10^{-12}\,kg}$. The mass distribution of the
interplanetary dust is best known near the Earth. It has been determined from lunar
crater counts and Earth-orbiting spacecraft (Pegasus, Explorer~16 and Explorer~23, Pioneer~8 and
Pioneer~9, LDEF and many others). Here we will use the mass distribution which
Divine %(1993)
\cite{divine-1993} derived from the measurements presented by Gr\"un et al.
\cite{gruen-et-al-1985}. %(1985).
In addition we
will use the radial dependence of the dust density that is based on zodiacal light
observations by Helios
\cite{leinert-et-al-1981}.
%(Leinert et al., 1981).

\subsection{Description of model distributions}

Several dynamical meteoroid models  have been developed in the last
decades to describe the interplanetary meteoroid environment. The most
comprehensive one so far is  the model of Divine  \cite{divine-1993}
and its extension by Staubach \cite{staubach-and-gruen-1995,
staubach-1996}  that synthesizes  meteor data, zodiacal light
observations, and in-situ measurements of  interplanetary dust. The
model describes dust concentrations and fluxes on the  basis of
meteoroid populations with distinct orbital characteristics. 
Gravitational (keplerian) dynamics and radiation pressure effects are
employed  in order to derive impact rates. Besides impact rates,
impact directions and  speeds can be modeled. The model includes dust
populations on bound orbits  around the Sun and an interstellar dust
population that penetrates the solar  system on unbound trajectories.

For the model description of interplanetary micrometeoroids we follow in principle the
method of Divine
\cite{divine-1993} %(1993)
with the extension introduced by Staubach
\cite{staubach-and-gruen-1995,staubach-1996}.
The spatial
dust density and the directional dust flux at each position in interplanetary space
are synthesized from 'model' distributions of orbital elements. Following Divine
%(1993)
\cite{divine-1993}
the interplanetary meteoroid complex is represented by several populations of dust
particles that are defined by their orbital elements and mass distributions. In order
to simplify the meteoroid model, several assumptions have been made: (1) the zodiacal
dust cloud is symmetric about the ecliptic plane, and (2) it has rotational symmetry
about the ecliptic polar axis. Therefore, in the model the spatial dust density depends
only on distance $r$ from the Sun and height $z$ above the ecliptic plane. Small
asymmetries relative to the ecliptic plane
%(cf. Leinert and Gr\"un, 1990)
\cite{leinert-and-gruen-1990}
or a small
offset of the cloud center from the Sun
%(Dermott et al., 1994)
\cite{dermott-et-al-1994}
are ignored. No time
dependences are considered, although for the sub-micrometer dust flux a 22-year cycle has
been suggested from theoretical considerations
%(Morfill and Gr\"un, 1979, Gustafson and Misconi, 1979, Hamilton et al., 1996)
\cite{morfill-and-gruen-1979-a,gustafson-and-misconi-1979,hamilton-et-al-1996}
and found in the data
\cite{landgraf-1998,landgraf-2000,landgraf-et-al-2000}.

In addition to solar gravity, orbits of micrometer-sized dust grains are affected by solar
radiation pressure. The ratio of radiation pressure force, $F_{\rm rad}$, to gravitational
attraction by the Sun, $F_{\rm grav}$, is defined by the factor
$ \beta = F_{\rm rad}/F_{\rm grav}$
\cite{burns-et-al-1979}.
%(Burns et al., 1979).
This $\beta$-value is strongly
dependent on material composition and structure of the dust particles. In the dynamical
dust model, $\beta$-values obtained by Gustafson (1994) from Mie calculations for
homogeneous spheres are assumed. For particles with masses $m > \rm 10^{-13}\,kg$
the influence of radiation pressure is negligible. At $\rm 10^{-14}\,kg$, $\rm 10^{-16}\,kg$,
and $\rm 5 \cdot 10^{-18}\,kg$ $\beta$-values of 0.3, 0.8, and 0.3, respectively, have
been assumed. These values represent a continuous dependence and reflect the
combination of the solar spectrum and particles size. 
For bound orbits $\rm \beta < 1$ is required.

In order to evaluate the flux at position $\bf r$ (bold characters indicate vector
quantities) the dust particle velocity, $\bf v$, has to be determined from the orbital
elements, perihelion distance, $r_1$, eccentricity, $e$, and inclination, $i$, and from the
assumed $\beta$-value. The relative velocity $\bf u_{\rm D}$ between a dust particle and
the spacecraft with velocity $\bf v_{\rm DB}$ is then $\bf u_{\rm D} = v - v_{\rm DB}$. 
The sensitivity of a detector can be expressed by its mass threshold $m_{\rm t}$ and its angular
sensitivity. For spinning spacecraft like Galileo and Ulysses the effective sensitive
area, $\rm \Gamma$, is a function of the angle, $\rm \gamma$, between the impact direction
and the spacecraft axis \cite{gruen-et-al-1992-a}. %(Gr\"un et al., 1992).

Populations of interplanetary meteoroids are described by independent distributions of
orbital inclination $ p_{\rm i} (i)$, eccentricity $p_e(e)$, perihelion distance $N_1(r_1)$, and
particles mass $H_{\rm M}(m)$, and the corresponding $\beta$-values. The detector threshold $m_{\rm t}$,
and the angular sensitivity $\Gamma$ are used as weighting factors for the calculated flux
\cite{divine-1993} \cite{gruen-et-al-1997-a}. For any assumed dust population (defined by its
distribution functions) and a given observation condition (spacecraft position,
$\bf r_{\rm DB}$, velocity, $\bf v_{\rm DB}$, detector orientation, $\bf r_{\rm D}$, and
sensitivity weighting factors), model fluxes and spatial densities can be calculated.

The distribution functions have been iterated by comparison with the corresponding
measurement: the difference between the measurement and the sum of model fluxes of all
dust populations is expressed by a residual. Each distribution function consists of a
limited number of parameters (e.g. the inclination and eccentricity distributions are
represented by seven parameters each) which were varied by an iteration algorithm in such
a way that the residual was minimized.

Populations of particles on heliocentric bound orbits have been defined by the procedure
described above. The populations are ad hoc populations used for fitting purposes
and do not correspond to specific sources of dust. 
Divine's core population (and the corresponding orbital distribution)
has been found to fit micrometeoroids of masses $m \rm > 10^{-13}\,kg$ that are not affected
by radiation pressure. Three populations of smaller meteoroids, with $\beta > 0$, on bound
heliocentric orbits fit the smaller particles detected by Galileo and Ulysses,
mostly inside 3 AU distance from the Sun. Although all populations are defined over
the whole mass range (${\rm  10^{-21}\,kg} \leq  m \leq {\rm 10^{-3}\,kg}$) each
population dominates in a narrow mass
interval for which a constant $\beta$-value has been assumed. The interstellar dust
population is well represented by Ulysses and Galileo dust measurements at large
heliocentric distances
\cite{gruen-et-al-1993,gruen-et-al-1994-b,baguhl-et-al-1995-b}.
The interstellar dust flow is, within the measurement uncertainties, aligned with the
flow of interstellar gas through the heliosphere \cite{witte-et-al-1993}. Observed dust
speeds are compatible with the $\rm 26\,km\,s^{-1}$ unperturbed gas speed. About 60\% of the
interstellar grains have masses between $\rm 10^{-17}\,kg$ and $\rm 10^{-15}\,kg$. With an
assumed value of $\beta = 1$, interstellar grains pass on straight trajectories through the
planetary system. This $\beta$ value is typical for the bulk of interstellar grains
observed by Ulysses (for a detailed discussion see 
\cite{landgraf-et-al-1999}). Model fluxes are calculated with these assumptions.

\begin{figure}[tbh]
\parbox{0.64\hsize}
{
\epsfxsize=7.8cm
\epsfbox{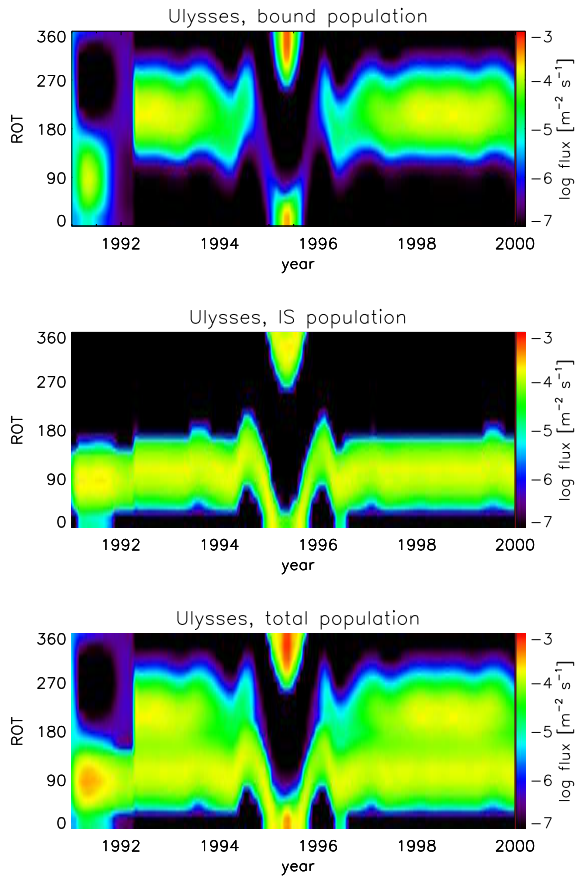}
}
\parbox{0.34\hsize}
{
        \caption{\label{Fig3_2_5}
Models of directional impact rates (of particles with masses $\rm m > 10^{-17}\,kg$)
onto Ulysses from launch to completion of its first out-of-ecliptic orbit. The color
coding of flux values is shown on the right side of each panel. Impact rates of
interplanetary dust on bound orbits (upper panel), interstellar dust (middle panel),
and the total impact rate (lower panel) are shown. The total impact rate can be
compared with the observed directional impact rates in Fig.~\ref{Fig3_2_2}
(the number of impacts -- diamonds and squares -- in a time and a rotation angle
interval corresponds to the modelled impact rate).
}
}
\end{figure}

Fig.~\ref{Fig3_2_5} shows the calculated directional fluxes for the considered time period.
The flux values are color coded and represent the flux in a phase space interval,
$\Delta$ ROT and $\Delta$ t. These model fluxes can be compared with the measurements shown in
Fig.~\ref{Fig3_2_2} where the number of impacts in the same phase space interval should be
considered.

\subsection{Comparison with zodiacal light}

The model can be used to derive the global interplanetary dust distribution as represented
by visible zodiacal light and infrared thermal emission observations. The latitudinal
density distribution was derived from zodiacal light observations
%(Leinert et al., 1981)
\cite{leinert-et-al-1981}
or from infrared observations with COBE
%(Reach et al., 1995).
\cite{reach-et-al-1995}.
Infrared brightnesses like those observed by COBE observations were
taken in a band approximately perpendicular to the solar direction and,
therefore, refer to dust outside Earth's orbit and do not
represent dust close to the Sun (this is in contrast to zodiacal light
observations which can be performed in almost any direction).
The difference between the
density distributions that fit the observed zodiacal and infrared brightnesses suggests
that zodiacal dust in the inner solar system has a wider distribution (to which the
zodiacal light measurements refer) than dust outside the Earth's orbit to which COBE
data refer. Divine's core population approximates the latitudinal density function of
zodiacal light observations for low latitudes ($\lambda < 10^{\circ}$) and that of
infrared observations for high latitudes ($\lambda > 20^{\circ}$). The density
distributions for the small particle populations have a wider latitudinal distribution
and are closer to the zodiacal light distribution. The assumed constant density of
interstellar dust provides a small but constant contribution at all latitudes.

\markright{JUPITER DUST STREAMS}

From the dust populations, Gr\"un et al. \cite{gruen-et-al-1997-a} have calculated model
brightnesses of the zodiacal light as observed from Earth. By far the
largest contribution comes from the core population with masses $ m >
10^{-13}\, {\rm kg}$. Particle populations with masses $ m < 10^{-13} {\rm  kg}$ contribute
less than 1\% to the zodiacal light brightness at 1~AU. In order to
obtain this result it was assumed that the albedo (ratio of visual
light reflected by a grain to the amount of incident light) is $ p = 0.05$
for the core population and $p = 0.02$ for the asteroidal population. We
had to assume that the small particle populations consist of very dark
particles (albedo of $p = 0.01$) in order for the sum of all populations
not to exceed the observed brightness.

\section{Electromagnetically interacting dust: Jupiter Dust Streams}
\label{jupiter_dust}

\markright{JUPITER DUST STREAMS}
\subsection{Observations}

\begin{figure}[tbh]
\vspace{-2.9cm}
\epsfxsize=10.3cm
\epsfbox[-120 255 450 805]{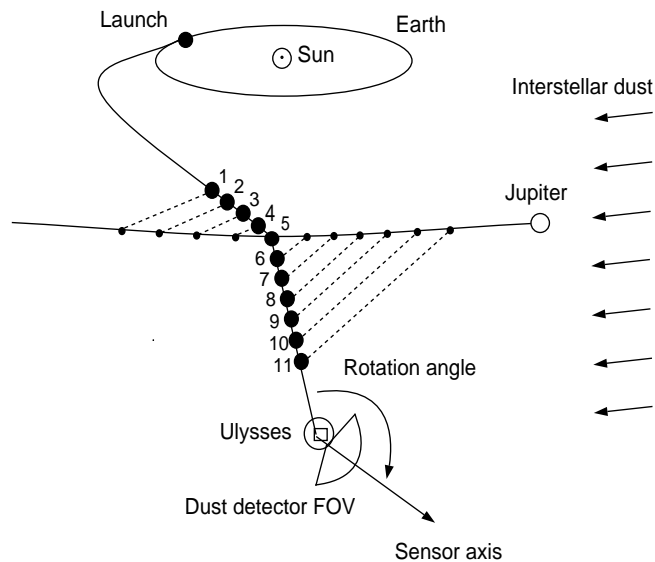}
\vspace{1.8cm}
        \caption{\label{geometry}
Ulysses trajectory and geometry of dust detection -- oblique
view from above the ecliptic plane. The position of the Sun and orbits of
Earth
and Jupiter (in the foreground) are also shown. The trajectory of
Ulysses after Jupiter closest approach is deflected into
an orbit inclined by $\rm 79^{\circ}$ to the ecliptic plane going
south. Numbers along the trajectory refer to positions of Ulysses
at which dust streams were detected, dotted lines point to Jupiter.
The spacecraft spins about an axis which points towards Earth.
The dust detector on board has a $\rm 140^{\circ}$ conical
field of view (FOV), and is mounted almost at a right angle
($85^{\circ}$) to the Ulysses spin-axis. Radiant directions
from which it can sense impacts therefore include the
plane perpendicular to the spacecraft-Earth line. The
rotation angle of the sensor axis at the time of a dust
impact is measured from the ecliptic north direction. Arrows
indicate the approach direction
of interstellar dust.
}
\end{figure}

\begin{figure}[ht]
\vspace{-3.5cm}
\epsfxsize=11.2cm
\epsfbox{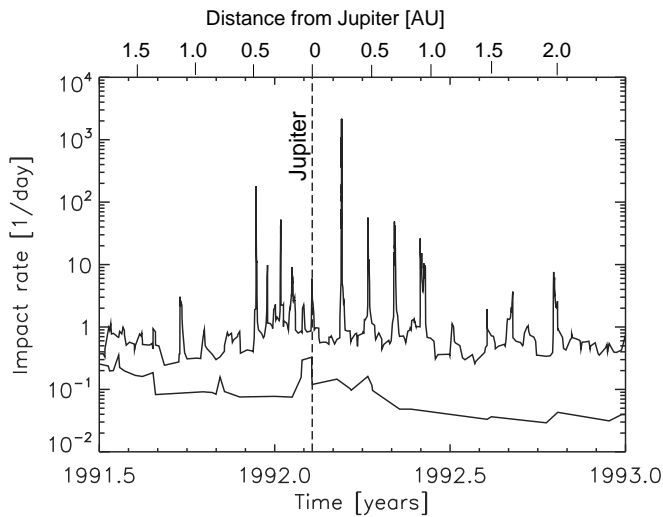}
\vspace{-4.5cm}
        \caption{\label{rate}
Impact rate of dust particles observed by Ulysses around Jupiter fly-by.
The curves show all impacts recorded (upper curve) and impacts of dust
particles with masses greater than $\rm 10^{-15}\,kg$. The impact rates
are means always including six impacts. The distance from Jupiter is 
indicated at the top.
}
\end{figure}

\begin{figure}[tbh]
\vspace{-2cm}
\epsfxsize=13.2cm
\epsfbox{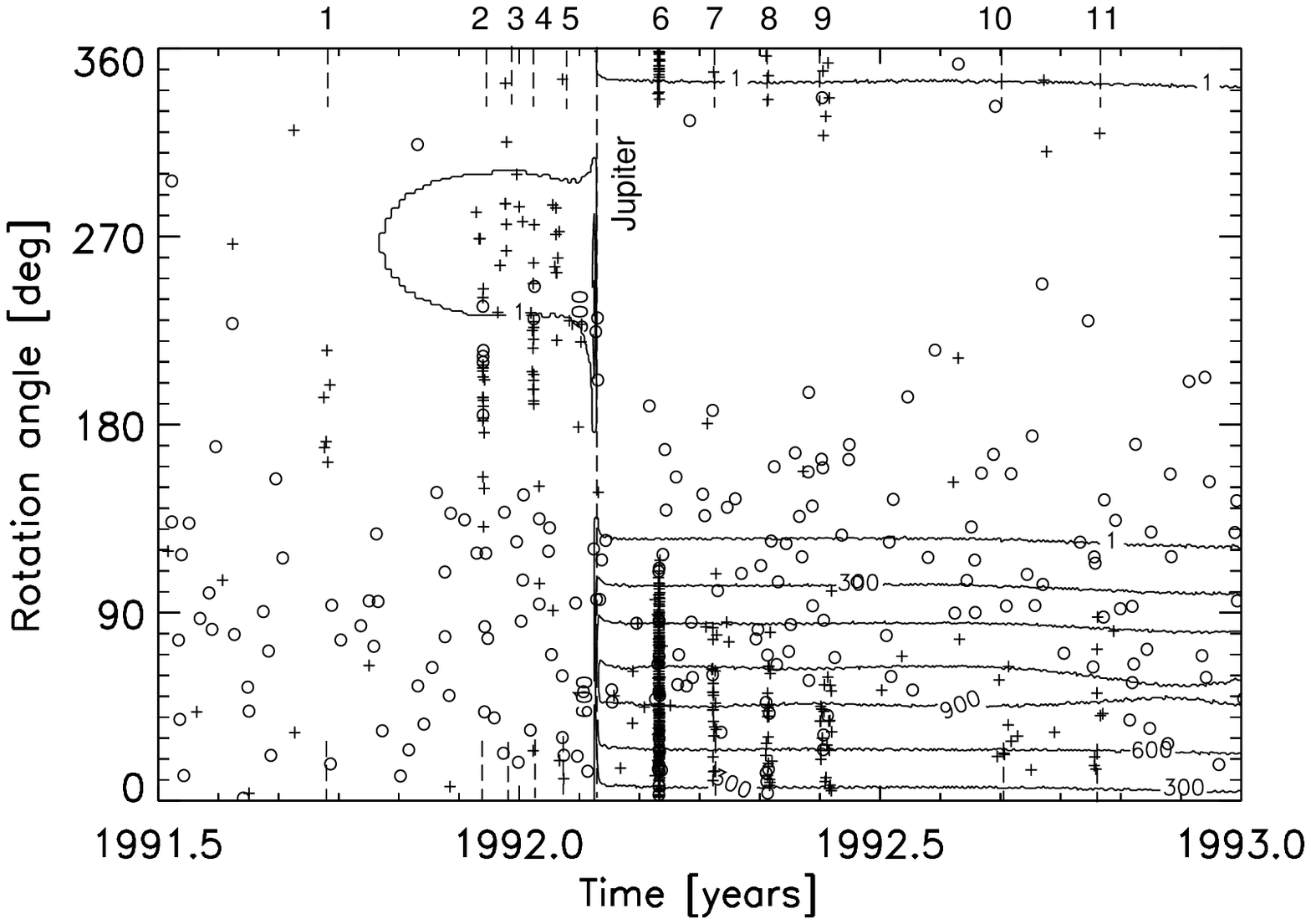}
\vspace{-8cm}
        \caption{\label{rot_angle}
Dust impact directions observed around Jupiter fly-by.
Crosses denote particles with masses $m < \rm 10^{-17}\,kg$,
circles denote particles with $m > \rm 10^{-17}\,kg$.
The dust streams are indicated by
dashed lines and are labelled 1 to 11. Closest approach to
Jupiter is indicated by a long dashed line. The contours
show the sensitive area of the dust sensor for particles
approaching on straight lines from the line-of-sight to Jupiter.
}
\end{figure}
On 10 March 1992, about one month after Jupiter fly-by, perhaps one of the
most unexpected findings of the Ulysses mission has been detected:
an intense stream of dust grains which was soon recognized to
originate in the Jupiter system \cite{baguhl-et-al-1993}. 
During this burst more than 350 impacts were detected within 26 hours.
This exceeded by more than two orders of magnitude the typical impact 
rates of interplanetary and interstellar dust seen before. Ten more
such dust streams could be revealed in the data when Ulysses
was within 2~AU from Jupiter
\cite{gruen-et-al-1993,baguhl-et-al-1993}.
The geometry of dust detection during Ulysses' Jupiter fly-by is shown in
Fig.~\ref{geometry} and the observed impact rate which shows the dust streams
as individual spikes is displayed in Fig.~\ref{rate}. The streams occurred
at approximately monthly intervals ($\rm 28 \pm 3$ days).
No periodic dust phenomenon in interplanetary space was known before
for small dust grains. Details of the streams are summarised in
Tab.~\ref{streams}.

The impact direction (rotation angle) of the dust stream particles is shown in
Fig.~\ref{rot_angle}.  The Jupiter system was the most likely source of the
streams because of the following reasons: 1) the streams were narrow and
collimated, which required a relatively close-by source. Otherwise they should
have been dispersed in space and time;
2) the streams were concentrated near Jupiter and the strongest
one was detected closest to the planet; 3) the streams
before Jupiter fly-by approached Ulysses from directions almost
opposite to the streams from after fly-by. All streams, however,
radiated from close to the line-of-sight direction to Jupiter.
4) The observed periodicity suggested that all streams originated
from a single source and seemed to rule out cometary or asteroidal
origins of individual streams. Especially, comet Shoemaker-Levy 9,
before its tidal disruption in 1992, was considered as a possible source
\cite{gruen-et-al-1994-a}.
Only two of the eleven streams, however, were compatible with an
origin from the comet.

\begin{center}
\begin{table}[ht]
\caption{\label{streams} Parameters of the Jupiter dust streams adopted from Baguhl et al.
\cite{baguhl-et-al-1993}.
The time corresponds to the center of the burst. Closest approach to Jupiter occurred on
92-039.5. Jupiter radius is $\rm R_J = 71,492\,km$.}

{\tiny

\begin{tabular}{lcccccc}
& & & & & & \\
\hline
& & & & & & \\[-2ex]
Stream & Days from &  Date (yr-day) & Number    & Mean        & Distance & Distance \\
number & Jupiter   & Duration (h)   & of        & rotation    & to       & to        \\
       & fly-by     &                & particles & angle (deg) & Sun (AU) & Jupiter(R$_{\rm J}$)\\
& & & & & &\\[-2ex]
\hline
& & & & & &\\[-2ex]
Stream & -136.7 & 91-267 & 6& 182& 4.30& 2356\\
\quad1 & & 76.0 & & & &\\
& & & & & &\\[-2ex]
Stream & -57.8 & 91-346 & 21 & 193 & 4.93 &  996\\
\quad2& & 10.4 & & & & \\
& & & & & &\\[-2ex]
Stream & -45.8 & 91-358 & 7 & 295 & 5.08 & 796 \\
\quad3 & & 18.0 & & & & \\
& & & & & &\\[-2ex]
Stream & -32.1 & 92-007 & 15 & 224 & 5.14 & 561 \\
\quad4 & & 10.5 & & & &\\
& & & & & &\\[-2ex]
Stream & -20.4 & 92-019 & 10& 259 & 5.29 & 367 \\
\quad5 & & 57.1 & & & & \\
 & & & & & &\\[-2ex]
Stream & 31.2 & 92-071 & 327 & 49 & 5.4 & 549 \\
\quad6 & & 34.3 & & & & \\
& & & & & & \\[-2ex]
Stream & 59.4 & 92-099 & 28 & 47 & 5.39 & 1017 \\
\quad7 & & 60.9 & & & & \\
& & & & & & \\[-2ex]
Stream & 87.4 & 92-126 & 33 & 29 & 5.38 & 1484 \\
\quad8 & & 55.6 & & & & \\
& & & & & & \\[-2ex]
Stream & 116.7 & 92-156 & 28 & 25 & 5.36 & 1965\\
\quad9 & & 107 & & & & \\
& & & & & &\\[-2ex]
Stream & 205.4 & 92-244 & 12 & 29 & 5.32 & 3418\\
\quad10 & & 215 &&&&\\
&&&&&&\\[-2ex]
Stream & 254.3 & 92-293 & 13 & 57 & 5.2 & 4200 \\
\quad11 & & 103.4 &&&&\\
&&&&&&\\[-2ex]
\hline

\end{tabular}

}

\end{table}
\end{center}

Confirmation of the Jupiter dust streams came from Galileo:
dust ``storms'' with up to 10,000 impacts per day were recorded  about
half a year before Galileo's arrival at Jupiter when the spacecraft was
within 0.5~AU from the planet \cite{gruen-et-al-1996}.
Later, with Galileo, the dust stream particles were detected within
Jupiter's magnetosphere
\cite{gruen-et-al-1997-b,gruen-et-al-1998}.
The data showed
a strong fluctuation with Jupiter's 10 hour rotation period which
demonstrated the electromagnetic interaction of the dust grains with the
ambient magnetic field of Jupiter. The dust streams seen
in interplanetary space are the continuation of the streams detected
by Galileo within the magnetosphere.

\subsection{Particle Masses and Speeds}

\begin{figure}[tbh]
\vspace{-2.0cm}
\epsfxsize=10.3cm
\epsfbox{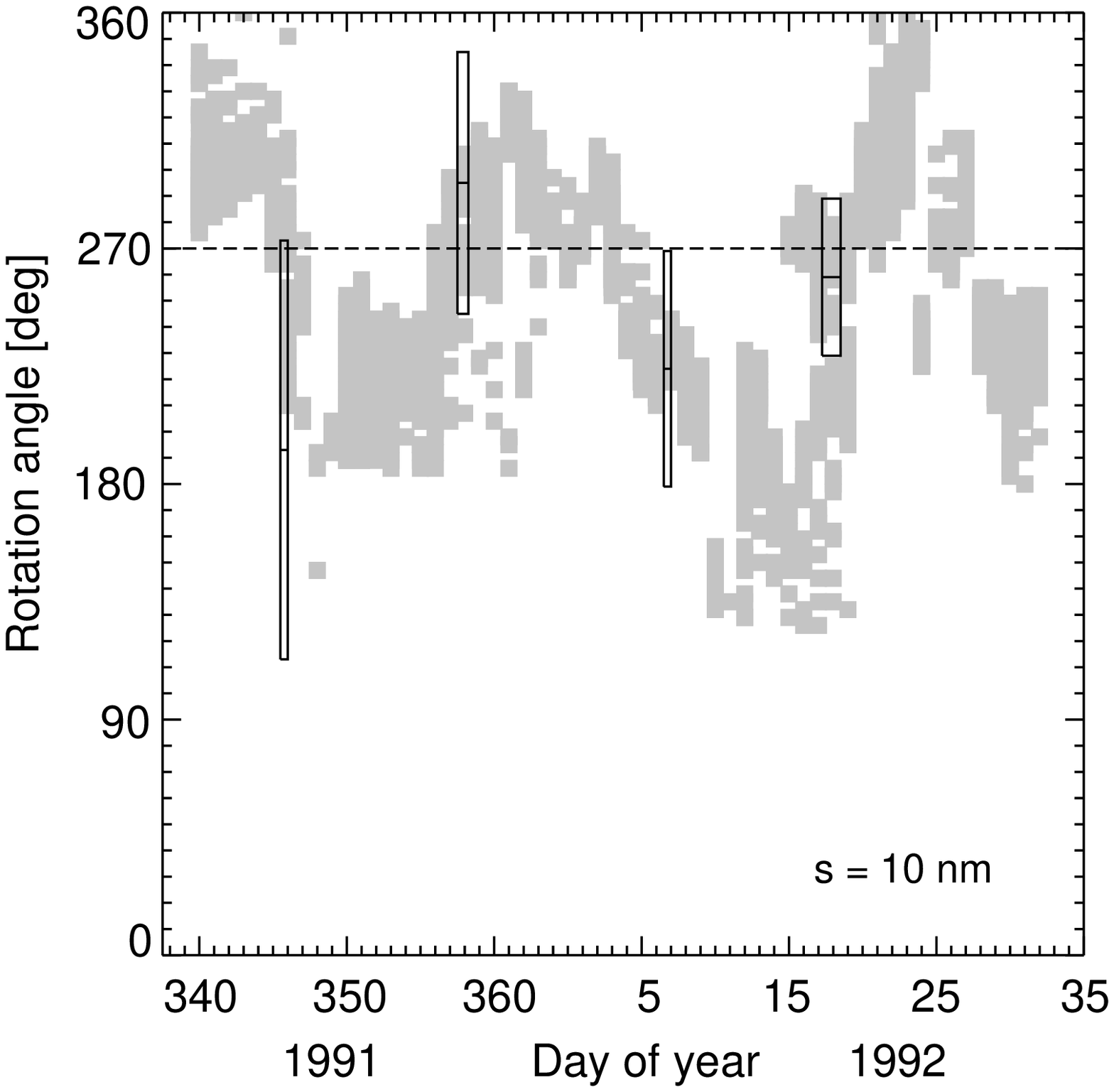}
\vspace{-1.7cm}
        \caption{\label{rot_angle_sim}
Simulated arrival directions (rotation angle) of 10\,nm dust particles from
Jupiter to Ulysses (shaded area) compared with directions from which
dust streams were observed by the Ulysses dust sensor (boxes; from
Zook, priv. comm.) The
box sizes indicate the duration and range of impact directions.
}
\end{figure}

The dust instruments on board Ulysses and Galileo have been
calibrated in the laboratory. The calibrated
range covers 3 to $\rm 70\,km\,s^{-1}$ in speed and $10^{-9}$
to $\rm 10^{-19}\, kg $ in mass
%(Gr\"un et al., 1995).
\cite{gruen-et-al-1995-a}.
By
applying these calibrations, masses and speeds of the stream particles
were $\rm 1 \cdot 10^{-19}\, kg$ to $\rm 9 \cdot 10^{-17}\, kg$ and
20 to $\rm 56\,km\,s^{-1}$, respectively. Uncertainties for the
velocity were a factor of 2, and, for the mass, a factor of 10.
Assuming an average density of $\rm 1\,g\, cm^{-3}$ the derived
sizes of the particles were 0.03 to $\rm 0.3\,\mu m$. These values,
however, were challenged by later investigations:

Although the particles' approach directions were close to the
line-of-sight to Jupiter, the approach direction of most
streams deviated too much from the direction to Jupiter to be explained
by gravitational forces alone. The deviation of the stream
direction from the Jupiter direction was correlated with the
magnitude and direction of the interplanetary magnetic field
(especially its tangential component). Strong non-gravitational
forces must have been acting on the grains to explain the observed
impact directions. In numerical simulations Zook et al. \cite{zook-et-al-1996}
integrated the trajectories of many particles backward in time and away from Ulysses
in various directions (Fig.~\ref{rot_angle_sim}).
Only particles which came close to Jupiter ($<\,100$ Jupiter radii)
were considered compatible with a jovian origin.
The dust particles were assumed to be electrically charged
to $\rm + 5\,V$
by a balance between solar photo-electron emission and
neutralization by solar wind electrons. The solar wind speed
was assumed to be $\rm 400\,km\,s^{-1}$. Forces acting on the
particles included Sun's and Jupiter's gravity and the interaction
with the interplanetary magnetic field as observed by Ulysses
\cite{balogh-et-al-1993}.
%(Balogh et al. 1993).

One important result from this analysis was that the
particles were about $10^3$ times less massive and 5 to 10 times
faster than the values implied by the calibration of the dust
instrument. Only particles with velocities in
excess of $\rm 200\,km\, s^{-1}$ and masses of the order of
$\rm 10^{-21}\,kg$ were compatible with an origin in the Jupiter
system. The corresponding particle radii were only 5 to $\rm 10 \,nm$.
Particles smaller than $\rm 5\,nm$ were unable to travel from
Jupiter to Ulysses. They were rather caught up by the solar wind and swept
away. Particles significantly larger than $\rm 25\,nm$ were not
compatible with the measured impact directions because they did not
interact strongly enough with the interplanetary magnetic field.
For 10\,nm sized particles the Lorentz force exceeds gravity by more
than a factor of 1,000 and the trajectories of such particles are
totally dominated by the interaction with the interplanetary
magnetic field. Although the simulated impact directions
can explain the observations, Ulysses should have detected particles
in between the dust streams when no impacts were detected. This
indicates that apart from
interaction with the interplanetary magnetic field there must be
other processes that modulate the particle trajectories.

The analysis by Zook et al. has demonstrated that the solar wind
magnetic field acts as a giant mass-velocity spectrometer for charged
dust grains. In particular, masses and impact speeds of the grains
derived from the instrument calibration
\cite{gruen-et-al-1995-a}
were
shown to be invalid for the tiny Jupiter dust stream particles.
Masses and speeds of stream particles derived from the Galileo
measurements agreed very well with the values obtained for the
streams detected by Ulysses (J.\,C. Liou and H. Zook, priv.
comm.).

\subsection{Dust Source and Particle Acceleration in Jupiter's Magnetosphere}

What is the source of the dust streams in the Jupiter system, and which
mechanism can accelerate the grains to sufficiently high speeds so that
they are ejected into interplanetary and, possibly, interstellar space? One
possible source was suggested even before the arrival of the two Voyager
spacecraft at Jupiter in
1979: the volcanoes on Io
\cite{johnson-et-al-1980,morfill-et-al-1980}.
Small grains entrained in volcanic plumes follow ballistic
trajectories and reach altitudes where they become exposed to the
Io plasma torus. When grains encounter this high-density plasma region,
they rapidly collect electrostatic charge and interact with the local
magnetic field. For grains smaller than $\rm 0.1\,\mu m$, the resulting
Lorentz force overcomes Io's gravity and these grains become
injected into Jupiter's magnetosphere. The injection velocity is
the relative velocity between the co-rotating magnetic field and Io,
i.\,e., $\rm 57\,km\,s^{-1}$.
This production mechanism sets an upper limit on the size of the dust
particles escaping from Io.

Dust grains which are positively charged and released at Io's distance
($r \rm  =  5.9\,R_J$, Jupiter radius $\rm R_J = 71,492\,km$) are
accelerated outward and leave the Jupiter system if their radii are
between about 9 nm to 180 nm
%(Gr\"un et al., 1998).
\cite{gruen-et-al-1998}. Smaller particles
remain tied to the magnetic field lines and gyrate around
them like ions do. Bigger grains move on gravitationally bound orbits
which are affected by the Lorentz force.
The size range of expelled particles narrows down if the source is
located closer to Jupiter, with the upper size limit staying approximately
constant at 180 nm. For example, at the outer distance of the gossamer
ring ($\rm 3 R_J$) the lower size limit is 28~nm
%(Hamilton and Burns, 1993).
\cite{hamilton-and-burns-1993}.
This radial variation of the lower size limit of expelled particles is
an indication of the source distance from Jupiter
%(Gr\"un et al., 1998).
\cite{gruen-et-al-1998}.
Sizes of ejected grains of about 10\,nm as determined by Zook et al.
\cite{zook-et-al-1996} are compatible with an Io source but contradict a source
significantly closer to Jupiter, like e.g. the gossamer ring.

\begin{figure}[tbh]
\epsfxsize=7.5cm
\epsfbox{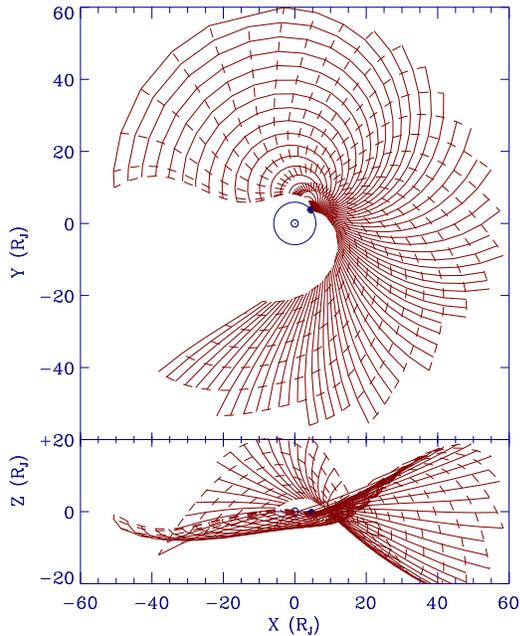}
        \caption{\label{sheet}
Jupiter's 'dusty ballerina skirt' formed by Jupiter stream particles moving
away from the planet in a warped dust sheet
\cite{gruen-et-al-1998}. Jupiter is in the center and the
particles are released from a point source at Io's distance.
        The dust sheet is
        defined by synchrones (solid lines, positions of dust
        particles released at the same time) and syndynes
        (dashed lines, positions of dust particles which have the same
        charge-to-mass ratio, $q/m$).  Adjacent synchrones are $1332\rm
        \,sec$ apart. Syndynes of differently
        sized particles range from $s = 9.0\rm\, nm$ to $s = 100\rm\, nm$, or,
        alternatively, at a surface potential of $U_g = +3\rm V$ to
        $q/m$-values between  984 and
        $8\rm\, C\,kg^{-1}$. Particle trajectories have been followed only out
to $\rm 50\,R_J$ which causes the ragged outer edge of the dust sheet. In
steady state the dust sheet extends to much larger distances. The
configuration of the dust sheet depends strongly on the phase of the
magnetic field.
}
\end{figure}

To a first order approximation, particles which are released from a
source close to Jupiter's equatorial plane (e.g. Io or the gossamer
ring) are accelerated along this plane.
Beside this outward acceleration, however, there is a significant out-of-plane
component of the electromagnetically induced force because Jupiter's
magnetic field is tilted by $9.6^{\circ}$ w.r.t. to the planet's rotation
axis. Thus, the dust particles are deflected out of Jupiter's equatorial
plane where they originated: Particles which are continuously released
from a point source move away from Jupiter in
a warped dust sheet, where particles are separated according to their
size (or more appropriately to their charge to mass ratio) and
time of generation. This scenario is depicted in Fig.~\ref{sheet}.
%The particle trajectories are typically bent out of the
%equatorial plane by 10 to $20^{\circ}$.
A detector in Jupiter's equatorial plane detects an increased number of
particles when this dust sheet
passes over its position. Fluctuations in the dust impact rate with 
5 and 10\,h periods and impact directions observed by Galileo
\cite{gruen-et-al-1997-b,gruen-et-al-1998}
can be explained by grains which are
electromagnetically coupled to the magnetic field. However, only
particles with a narrow size range of about $\rm 10\,nm$ can explain the
observed features. Modelling implies that larger or smaller particles
have not been detected by the Galileo dust instrument.
The 28 day periodicity detected by Ulysses
in interplanetary space (cf. Fig~\ref{rate}) can be explained by a
resonance between the
rotation period of Jupiter (and, hence, its magnetic field) and Io's
orbital period
\cite{horanyi-et-al-1993-a}.

%An alternative model was proposed by Hamilton and Burns
%\cite{hamilton-and-burns-1993}.
%These authors suggested Jupiter's gossamer ring as a source
%of the dust streams. Their model explained the observed
%periodicity with the solar rotation period and reproduced
%the  asymmetry (Tab.~\ref{streams}) in the number of streams before
%and after Jupiter fly-by as well as the strength of individual streams.

\begin{figure}[tbh]
\epsfxsize=12.2cm
\vspace{-3.4cm}
\epsfbox{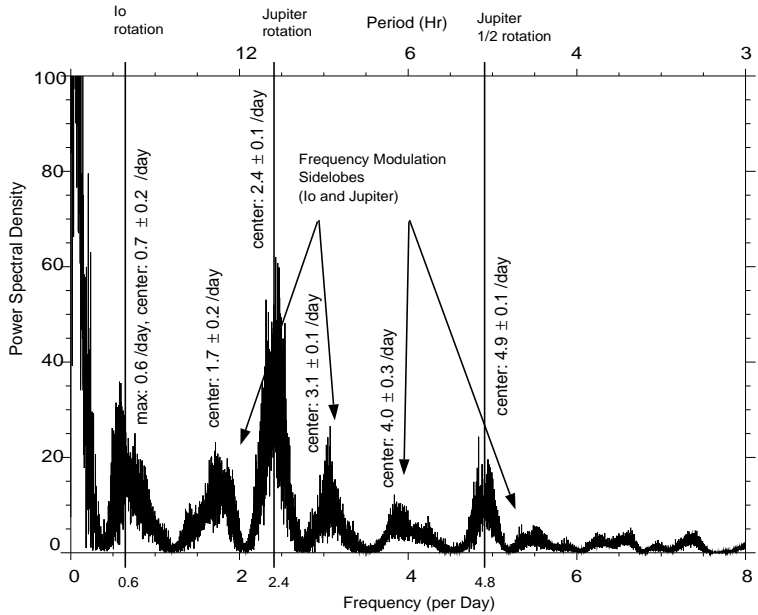}
\vspace{-4.2cm}
\caption{\label{periodogram}
A Scargle-Lomb periodogram \cite{scargle-1982} for two years of
Galileo dust data (1996 and 1997; from \cite{graps-et-al-2000}).
A source with a 10\,h period, Jupiter's magnetic
field, is modulated by Io's frequency which leads to peaks at 13.9\, and
7.7\,h.  There are additional frequencies, a
half period of Jupiter's magnetic field (4.9\,h), Io as a single source
(41.9\,h), and a
low-frequency oscillation trend due to the spacecraft orbital
geometry.
}
\end{figure}

Although modelling results are compatible with Io being the
source of the dust stream particles, modelling alone could not
prove the source.
%Time series analysis on a small data set early in the mission did
%not show a clear indication of Io's orbital period in the impact
%rate (Gr\"uen et al. 1998).
In addition to the 5h and 10h periods which are compatible with
Jupiter's rotation period, a modulation of the impact
rate with Io's orbital period (42\,h) was found in the data
\cite{krueger-et-al-1999-a}. A periodogram which transformed
two years of impact rate data into frequency space shows an amplitude
modulation of Jupiter's magnetic field frequency and Io's orbital
frequency (Fig.~\ref{periodogram}; \cite{graps-et-al-2000}). The periods
of the magnetic field and
Io modulate each other, and the only explanation is Io being the source
of the ``carrier frequency''.
Photometric observations of the Io plumes obtained with Voyager
imply a size range of 5 to $\rm 15\,nm$ \cite{collins-1981} and recent
Hubble Space Telescope observations constrained the particles
to be smaller than 80 nm
\cite{spencer-et-al-1997}.
This is compatible with the results of Zook et al.
\cite{zook-et-al-1996}.

%What did we see in frequency space? We believe that
%we see an amplitude modulation of Jupiter's magnetic field frequency
%and Io's rotational frequency. That is, that the periods of the field
%and Io and modulate each other, and the only explanation is Io being
%the source or ``carrier frequency.''

The characteristic signature of an amplitude modulation of oscillation
frequencies is a carrier frequency ($\omega_0$) with side frequencies,
or ``modulation products,'' ( $\omega_1+\omega_0$) and
($\omega_1-\omega_0$). If Io's orbital frequency is a carrier
frequency, and charged dust trapped in Jupiter's magnetic field is a
frequency of Jupiter's rotation, that is amplitude-modulating Io's
signal, then one would see a frequency at Jupiter's rotation period
with side frequencies plus and minus Io's orbital frequency. The
frequency-transformed data obtained with Galileo directly display
this amplitude-modulation scenario.

The dynamics of the charged nanometer-sized dust particles which are
released from Io, pass the Io torus and are finally ejected out of
Jupiter's magnetosphere is strongly affected by the geometry of
the magnetic field. In a completely radially symmetric magnetic field
configuration no 10\,h period should be evident in the impact rate,
only the 5\,h period should be seen. The prominent modulation
of the rate with the 10\,h period indicates that there must be
a variation in the acceleration of the particles which is correlated with
jovian local time. The Io dust stream particles will serve as
test particles for future investigations of the jovian magnetic
field environment.

\section{Interstellar Dust}          \label{interstellar_dust}

Prior to the Ulysses dust measurements, the abundance and even the
mere existence of interstellar dust grains inside the solar system
was controversial. Interstellar grains
which are not bound to the Sun or any other star, but belong to the
interstellar medium of our Galaxy, should pass by the Sun on its
way through that medium. By using the density of
the interstellar medium surrounding the solar system and the general
assumption that $1\%$ of the mass of the interstellar medium is
contained in dust grains \cite{holzer-1989}, it was postulated that
the flux
%(particles per unit area and time)
of these grains in the solar system has to be of the order of
$10^{-3}\;{\rm m}^{-2}\;{\rm s}^{-1}$. This is higher than the total
flux of cosmic dust observed at Earth orbit
\cite{gruen-et-al-1985}. This non-detection lead to the conclusion
that there has to be a physical process which removes interstellar
dust from at least the inner solar system.  Levy and Jokipii
\cite{levy-et-al-1976} showed that the dominant force on small dust
grains in the solar system is the Lorentz force caused by the solar
wind magnetic field sweeping by the electrically charged dust
grains. Electron emission due to solar UV photoeffect, in equilibrium
with the sticking of electrons from the surrounding plasma
environment, charges the grains to a positive surface potential. Since
the Lorentz force acts perpendicular to the solar wind velocity
vector, this seemed to be a mechanism for diverting the stream of
interstellar dust grains out of the solar system. Nonetheless, it was
argued \cite{gustafson-and-misconi-1979,morfill-and-gruen-1979-b} that
the Lorentz force not only diverts, but also concentrates interstellar
dust, depending on the phase of the solar cycle. Furthermore, the
Lorentz force is not as effective on larger grains as it is on small
ones, because it accelerates grains proportional to their
charge-to-mass ratio $q/m$. The charge-to-mass ratio, in turn, is
inversely proportional to the grain radius $s$, if a constant surface
potential and a spherical shape of the grain are assumed.

The size range $0.005\;{\rm \mu m} \leq s \leq 0.25\;{\rm \mu m}$ of
classical interstellar dust grains, which have been postulated to
explain the extinction of starlight \cite{mathis-et-al-1977}, indicated that
these grains should be heavily affected by interaction with the solar
wind magnetic field. Before Ulysses the proposed properties of interstellar
dust were
\begin{itemize}
\item[(a)] there are no big ($s > 0.25\;{\rm \mu m}$) dust grains in
the diffuse interstellar medium surrounding the Sun, and
\item[(b)] the small ones are prevented from entering the planetary
system by the solar wind magnetic field.
\end{itemize}

\subsection{Discovery and Identification of Interstellar Dust Grains}
After Ulysses had flown by Jupiter in February 1992, dust
impacts were detected predominantly from a rotation angle between
$0^\circ$ and $180^\circ$. As can be seen in Figure~\ref{geometry},
this corresponds to grains approaching from the
retrograde direction. Significantly fewer impacts were detected from the
prograde direction from which grains released from solar system bodies
are expected, (cf Figure~\ref{Fig3_2_2}).
Attempts to explain the measurements by a population of retrograde
interplanetary grains failed because the impact rate of grains from
the retrograde direction was constant while Ulysses was moving to
higher and higher heliocentric latitudes \cite{baguhl-et-al-1996}. Since the
impact rotation angle of about $100^\circ$ coincided with the upstream
direction of gas from the local interstellar medium \cite{witte-et-al-1993}, it
was concluded that Ulysses detected particles belonging to a stream of
dust grains originating from the interstellar medium \cite{gruen-et-al-1993}.
In this respect the delay of the Ulysses mission and the resulting
change in the final orbit turned out to be fortunate. On the
initially planned orbit the Ulysses spin-axis would have pointed
nearly parallel to the upstream direction of the interstellar dust most
of the time. Since the dust detector points nearly
perpendicular to the spin-axis, the impact rate of interstellar dust grains
would have been strongly reduced and the identification of
interstellar impacts would have been much more difficult.

To support the evidence for interstellar grains in the solar
system, the impact velocities of grains detected from the retrograde direction
after Jupiter fly-by have been analyzed. Most measured impact velocities, although
uncertain by up to a factor of $2$, indicated that the grains had
heliocentric velocities in excess of the local solar system escape
velocity, even if radiation pressure effects were neglected
\cite{gruen-et-al-1994-b}. Analysis of the data collected with the
Galileo dust instrument further supported the
interpretation of the Ulysses data as impacts of interstellar
grains \cite{baguhl-et-al-1995-b}. The discovery of interstellar grains in
the solar system contradicted the considerations of the removal
of interstellar dust summarized in points (a) and (b)
above. This contradiction is resolved by the following considerations
\begin{itemize}
\item[(A)] interstellar grains with radii larger than $0.25\;{\rm \mu
m}$ exist in the diffuse medium. They cannot be detected by
measuring the extinction of starlight because they scatter and absorb
light independently of the wavelength of the incident radiation.
Therefore they do not influence the shape of the interstellar extinction curve, and
\item[(B)] the solar wind magnetic field can indeed
concentrate interstellar grains in the solar system. This depends on
the phase of the solar cycle
\cite{gustafson-and-misconi-1979}.
\end{itemize}

After it was settled that the data set gathered by the Ulysses dust
detector contains interstellar grains and that they dominate the
impacts detected in the outer solar system, the question was raised
how we can distinguish them from impacts of interplanetary grains. In
other words, how can we find a subset of the Ulysses dust data which
contains only impacts from interstellar grains with high confidence?

As described above, the impact direction is the major
characteristic that distinguishes interstellar from interplanetary
grains in the outer solar system. Therefore, the identification
criterion for interstellar impacts in the Ulysses data set is:
\begin{flushleft}
\em Impacts that have been detected (a) after Ulysses left the ecliptic
plane, and (b) with a rotation angle for
which the interstellar gas upstream direction lies within the field of
view of the dust detector (plus a $10^\circ$ margin) are preliminarily
identified as interstellar impacts.
\end{flushleft}
The so defined data set still contained a contribution from
interplanetary impacts because during the short time of the first
perihelion passage of Ulysses around March 1995,
interplanetary grains hit the detector from the same direction as
interstellar grains. Furthermore, Hamilton et
al. \cite{hamilton-et-al-1996} suggested that very small ($s < 0.1\;{\rm \mu
m}$) interplanetary grains can reach high latitudes when they are
electromagnetically ejected from the solar system during favorable
phases of the solar cycle. These grains can also contribute to the data set
defined above. Therefore, impacts have been removed from the
preliminary data set that a)
have been measured around ecliptic plane crossing
($\pm 60^\circ$ ecliptic latitude), or b) that
have been measured above the solar poles with very small
impact signals ($Q_{\rm I} < 10^{-13}\;{\rm C}$), indicating small masses.

In Fig.~\ref{Fig3_2_2} we show all dust impacts detected by
Ulysses in a rotation-angle-vs-time diagram. The majority of particles in the
intermediate mass range is compatible with an interstellar origin. They
dominate in the outer solar system.

\subsection{Mass Distribution and Cosmic Abundances}

As explained above, the existence of interstellar grains in the solar
system can be explained by the fact that they are too big to be swept
away by the solar wind magnetic field. Why can the dust models based
on extinction measurements not simply be extended to account for big
grains? The reason is that the mass of solid matter in the
interstellar medium is limited by the {\em cosmic abundance of heavy
elements}. The most abundant elements in the Galaxy are hydrogen and
helium. In dust grains the amount of both of these elements is
negligible as compared to heavier elements. The mass contained in
solid dust grains is therefore limited by the available mass contained
in elements heavier than helium. On average, $99\%$ of the mass of the
local interstellar medium is carried by hydrogen and helium atoms
\cite{holzer-1989}. Therefore, the total mass of dust grains cannot
exceed $1\%$ of the overall mass of the medium.

If we calculate a mass distribution of the dust grains,
the contribution of each grain mass
interval to the total dust mass can be determined. The mass
distribution for previous interstellar grain models postulates that
most mass is contained in the large grains, although they are less
abundant. If this grain mass distribution were extrapolated to
infinitely high masses, the total mass contained in dust grains would be
infinite. This would contradict cosmic abundance measurements.
Therefore, the big interstellar grains found by Ulysses are important
for understanding the composition of the interstellar
medium.

\begin{figure}[tbh]
\parbox{0.49\hsize}
{
\epsfxsize8cm
\epsfbox{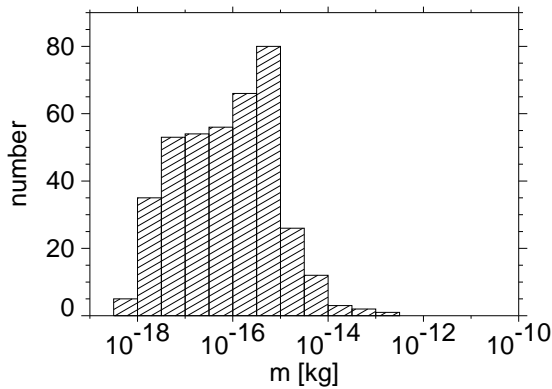}
}
\parbox{0.5\hsize}
{
\caption{\label{fig_uls_masshist} Histogram of the mass distribution
of interstellar grains detected by the Ulysses dust instrument. The 
detection threshold for grains impacting with $\rm 20\,km\,s^{-1}$ is
$\rm \sim 10^{-18}\,kg$.
}
}
\end{figure}

We can quantify the contribution of the grains measured by
Ulysses to the overall mass in dust in the interstellar medium by
calculating the mass distribution for the Ulysses measurements and comparing
it with the mass distribution of existing interstellar dust models.
The mass distribution of interstellar grains measured
by Ulysses is shown in Figure~\ref{fig_uls_masshist}.
The detected masses range from $10^{-18}$ to $10^{-13}\;{\rm
kg}$ and most grains have masses between $3\cdot 10^{-16}$
and $1\cdot 10^{-15}\;{\rm kg}$ \cite{landgraf-et-al-2000}.

\begin{figure}[tb]
\epsfxsize=\hsize
\epsfbox{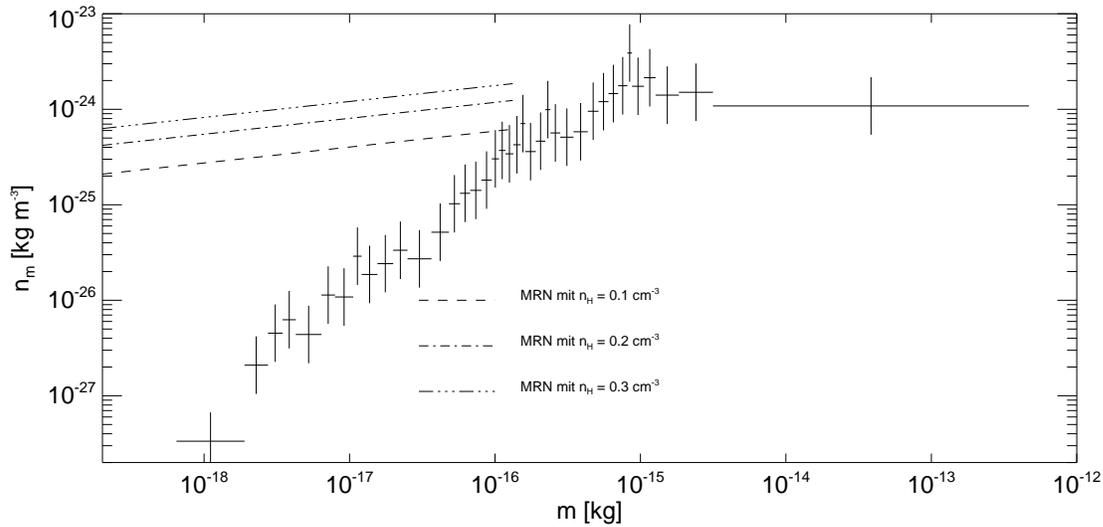}
\caption{\label{fig_massdens} Mass density per logarithmic mass
interval \protect\cite{frisch-et-al-1999}. Crosses show the distribution of
grains detected by Ulysses and Galileo. The broken lines represent the
mass per logarithmic mass interval for the MRN distribution assuming three
different overall densities of the local interstellar medium (measured
by number $n_{\rm H}$ of hydrogen atoms per unit volume). The upper
cut-off of the MRN distribution is $2\cdot 10^{-16}\;{\rm kg}$.}
\end{figure}

To compare the in-situ measurements with the mass distribution of dust
models based on extinction measurements (\cite{mathis-et-al-1977}, 
MRN distribution hereafter), we calculate
the contribution of each logarithmic mass interval to the overall mass
density in the interstellar medium. The result is shown in Figure
\ref{fig_massdens}.

Comparing the in-situ measurements with the MRN distribution one
finds two discrepancies:
\begin{itemize}
\item Small grains ($m<10^{-16}\;{\rm kg}$) are deficient.
\item The in-situ distribution extends to much larger masses than the
MRN cut-off.
\end{itemize}
The first discrepancy was explained to be caused by the partial removal
of small grains due to their interaction
with the solar wind magnetic field
\cite{gruen-et-al-1994-b,landgraf-2000}. In addition, interstellar
grains with diameters around $0.4\:{\rm \mu m}$ are deflected by solar
radiation pressure and are thus deficient in the inner solar
system \cite{landgraf-et-al-1999}.
However, as explained above, the
existence of {\em large} interstellar grains cannot be accounted for by
dust models based on extinction measurements. Figure
\ref{fig_massdens} shows that the large grains detected by Ulysses and
Galileo contribute significantly to the overall mass density of
heavy elements (heavier than helium) in the local interstellar
environment. Calculating the overall mass of the local interstellar
medium locked up in grains and assuming that also smaller grains
contribute, it was found \cite{frisch-et-al-1999} that dust grains make up
$2\%$ of the mass of interstellar matter surrounding the solar
system. This is twice the amount of dust expected in the average
medium of the Galaxy, which is not expected, since the local medium
is less dense than average and already contains large amounts of heavy
elements in the gas phase \cite{frisch-et-al-1999}.

\section{Summary and conclusions}

The prime goal of the Ulysses mission with respect to dust was to
determine the 3-dimensional structure of the interplanetary dust
cloud. This goal has been fully accomplished. Ulysses dust data is a
crucial ingredient of the comprehensive dynamical model that describes
the interplanetary dust cloud \cite{gruen-et-al-1997-a}.

Three more dust populations have been discovered that have not been seen
before by any other in-situ dust experiment: jovian dust stream
particles near Jupiter, electromagnetically controlled $\beta$-meteoroids
above the poles of the Sun, and interstellar grains everywhere else.

The jovian dust streams discovered by Ulysses demonstrated the fact that
dust may become an intimate player in magnetospheric processes. We are
still at the beginning of our understanding of these interrelations but
already now the importance of dust becomes evident. The dust streams are
monitors of the volcanic plume activity on Io in a way that is not
accomplished yet by any other observational method. So far, only
occasional images from Voyagers and Galileo have shown the plumes which
are not necessarily related to lava outpours on Io also observed from Earth.
Electromagnetically coupled dust grains are probes of the plasma
environment in the Io torus where they get their initial charge before
they are emitted by Jupiter's magnetic field from the jovian system.
Dust stream particles observed at different times originate from
different portions of the Io torus. Therefore, observations tracing
back to different local times in the Io torus may reveal local time
variations in the torus that are predicted by theory. Tiny dust grains
can transport material to regions of the magnetosphere where other known
magnetospheric processes are unable to transport ions to. Glows of
sodium and other elements may be examples for such a transport. Finally,
dust stream particles will be used as carriers of information about the
chemical composition in Io's volcanic plumes when Cassini flys by Jupiter
in December 2000. The dust instrument on board may analyze their
chemical composition.
In February 2004 Ulysses will fly by Jupiter within a distance of 0.8 AU.
A repetition of dust stream measurements in interplanetary space
will be beneficial to test our understanding of this new phenomenon.

Yet another population of very small dust particles was observed at high
ecliptic latitudes over the Sun's pole. This population has been
suggested to be electromagnetically deflected $\beta$-meteoroids
generated at low ecliptic latitudes and subsequently deflected by the
contemporary interplanetary magnetic field to high latitudes. Long-term
observations (over two solar cycles) of these particles may prove the
prediction that during one solar cycle these particles are found to
leave the solar system at high latitudes again while during the other
cycle they escape near the ecliptic. This way, these particles are the
Sun's analog to Jupiter's dust streams. The study of these particles
will provide one further piece in the puzzle of how much dust and from
which region in the solar system solid particles are ejected to the interstellar
medium.

The question if the interstellar dust grains measured in the solar
system by Ulysses and Galileo are ``native'' to the local interstellar
medium, i.e. if they originate from the medium, or if they have been
injected, is connected to their kinematic state with respect to the medium.
Can we measure a systematic motion of the grains relative to the gas?

\begin{figure}[ht]
\vspace{0.5cm}
\epsfbox{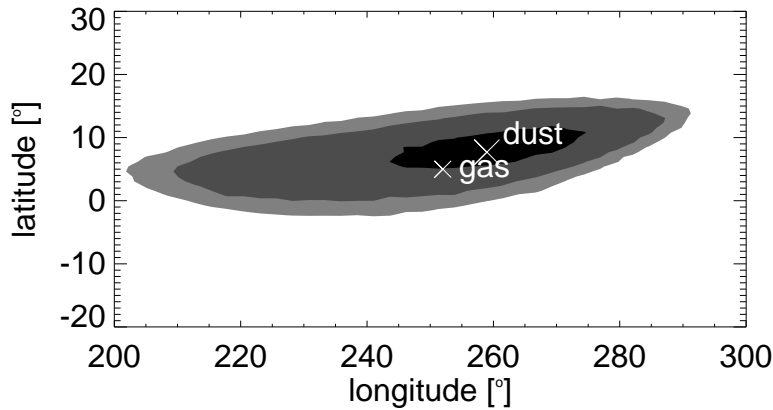}
\vspace{1cm}
\caption{\label{fig_upstream_dir} Ecliptic longitude and latitude of
the upstream direction of interstellar dust grains measured
by Ulysses after Jupiter fly-by. The contour plot of a $\chi^2$ analysis
shows $1\sigma$, $2\sigma$, and $3\sigma$
confidence levels in black, dark grey, and light grey, 
respectively. The minimum $\chi^2$ is achieved at $259^\circ$
longitude and $8^\circ$ latitude (indicated by the cross labeled
``dust''). The measurements of the helium upstream direction are
indicated by the smaller cross labeled ``gas''.}
\end{figure}

The velocity vector of neutral helium atoms with respect to the Sun
has been determined with the Ulysses GAS experiment. The
helium upstream direction was found to be $252^\circ$ ecliptic
longitude and $5^\circ$ latitude, and the absolute
relative velocity is $25.22\;{\rm km}\;{\rm s}^{-1}$
\cite{witte-et-al-1993}. By determining the dust upstream direction and
comparing it to the values for the helium atoms, we can establish the
kinematic relationship between dust and gas. Since the Ulysses dust
detector has a field of view of $\pm 140^\circ$, the impact direction
of an individual grain is not accurately determined and we have to
apply a statistical analysis of the distribution of measured rotation
angles. Figure \ref{fig_upstream_dir} shows the result.

We find that the dust upstream direction coincides with the gas
direction on the $1\sigma$ confidence level. This indicates, that
interstellar gas and dust is in rest with respect to
each other. However, since the determination of the dust upstream
direction has considerable uncertainties, a velocity dispersion as
large as $10\;{\rm km}\;{\rm s}^{-1}$ between dust and gas cannot be
ruled out \cite{landgraf-1998}. The Ulysses data
indicate a dust component at rest with the local interstellar gas.

For interstellar dust it seems that Ulysses has raised a number of new
questions by solving the question about the existence of interstellar
dust grains in the solar system. These can by summarized as follows:
\begin{itemize}
\item Is the concept of cosmic elementary abundances correct on all
length scales or are there small scale inhomogeneities of the galactic
chemistry?
\item If the answer to the first question is negative, the interstellar
grains we measure with Ulysses cannot originate from the interstellar
gas of the surrounding medium. So how did they get there and what
governs the dynamics of solid grains in the interstellar medium of our Galaxy?
\end{itemize}

To further confirm our findings and to improve the understanding of
the interstellar medium surrounding our solar system, long term
measurements are needed because the total flux of interstellar dust
grains is low. Ulysses provides an ideal platform for these measurements,
because of its fortunate orbit geometry.

\vspace{1cm}

{\bf Acknowledgements.}
We thank the ESA/NASA Ulysses projects for effective and successful
mission operations.
This work has been supported by Deutsches
Zentrum f\"ur Luft- und Raumfahrt e.V. (DLR) and by the Sonderforschungsbereich
Sternentstehung of the Deutsche Forschungsgemeinschaft (DFG).

%\vfill

\end{document}